\documentstyle{article}
\renewcommand{\theequation}{\thesection.\arabic{equation}}
\newcounter{hran} \renewcommand{\thehran}{\thesection.\arabic{hran}}

\def\bmini{\setcounter{hran}{\value{equation}}
  \refstepcounter{hran}\setcounter{equation}{0}
  \renewcommand{\theequation}{\thehran\alph{equation}}\begin{eqnarray}}

\def\bminiG#1{\setcounter{hran}{\value{equation}}
\refstepcounter{hran}\setcounter{equation}{-1}
\renewcommand{\theequation}{\thehran\alph{equation}}
\refstepcounter{equation}\label{#1}\begin{eqnarray}}

\def\emini{\end{eqnarray}\relax\setcounter{equation
}{\value{hran}}\renewcommand{\theequation}{\thesection.\arabic{equation}}}

\def\beq{\begin{equation}}
\def\eeq{\end{equation}}
\newcommand{\be}{\begin{equation}}
\newcommand{\ee}{\end{equation}}
\def\bea{\begin{eqnarray}}
\def\eea{\end{eqnarray}}

\newcommand{\barre}[1]{%
	\setbox1=\hbox{$#1$} \dimen2=\ht1 \dimen3=\dp1 \dimen4=\wd1
	\setbox2=\hbox{\sl /}
	\dimen1=\wd1 \advance\dimen1 by -\wd2 \divide\dimen1 by 2
	\advance\dimen1 by \wd2 \advance\dimen1 by 0.4pt
	\setbox3=\hbox to \wd1{\hss \box1 \kern -\dimen1 \box2\hss}
	\ht3=\dimen2 \dp3=\dimen3 \wd3=\dimen4
	\box3
	}

\def\1{{\rm 1 \kern -.10cm I \kern .14cm}} \def\R{{\rm R \kern -.28cm I
\kern .19cm}}
\begin{document}

\begin{titlepage}
\begin{flushright}    UFIFT-HEP-98-10 \\ 
%hep-ph/9805237 
\end{flushright}
\vskip 1cm
\centerline{\LARGE{\bf {Dynamical supersymmetry breaking }}}
\vskip .5cm 
\centerline{\LARGE{\bf {in a superstring inspired model}}}
\vskip 1.5cm
\centerline{\bf Nikolaos Irges \footnote{Supported in part by the
United States Department of Energy under grant DE-FG02-97ER41029.}} 
 
\vskip .5cm
\centerline{\em  Institute for Fundamental Theory,}
\centerline{\em Department of Physics, University of Florida}
\centerline{\em Gainesville FL 32611, USA}
\vskip 1.5cm

\centerline{\bf {Abstract}}
\vskip .5cm
We present a dilaton dominated scenario for 
supersymmetry breaking in a recently constructed
realistic superstring inspired model with 
an anomalous $U(1)$ symmetry. Supersymmetry is 
broken via gaugino condensation due to a confining $SU(N_c)$ gauge group 
in the hidden sector. 
In particular, we find that by imposing on the model the phenomenological constraint of the
absence of observed flavor changing neutral currents, 
there is a range of parameters related to the hidden sector and the  
K$\ddot{\rm a}$hler potential for which we obtain a low energy
spectrum consistent with present experimental bounds.   
As an illustrative example, 
we derive the low energy spectrum of a specific model. We find that the 
LSP is the lightest neutralino with a mass of $53\; GeV$ and the lightest
Higgs $h^0$ has a mass of $104\; GeV$. 

\vfill
\begin{flushleft}
May 1998 \\
\end{flushleft}
\end{titlepage}

%%%%%%%%%%%%%%%%%%%%%%%%%%%%%%%%%%%%%%%%%%%%%%%%%%%%%%%%%%%%%%%%%%%%%%%%%%%%%%%%%%%%%%%%%%%%%%%%%%%%%%%%%%%%%%%%%

%%%%%%%%%%%%%%%%%%%%%%%%%%%%%%%%%%%%%%%%
\section{Introduction}
%%%%%%%%%%%%%%%%%%%%%%%%%%%%%%%%%%%%%%%%

During the last few years there has been an increasing activity in 
trying to construct a complete and  phenomenologically 
viable model with an anomalous $U(1)$ gauge symmetry-$X$-,  
remnant of superstring compactification,
with its anomalies canceled via the Green-Schwarz
mechanism \cite{GS}. 
The Green-Schwarz anomaly canelation mechanism occurs if the non
zero anomaly coefficients $C_j$ and the corresponding Kac-Moody levels $k_j$, satisfy
\be {C_j\over k_j}=16\pi ^2{\delta }_{GS}, \;\; {\rm for
\; all }\; j, \ee
with
\be {\delta }_{GS}={C_g\over {192\pi ^2}}, \;\;\;\; C_g=Tr(X).\ee  
In these models, 
the vacuum expectation value of the dilaton generates a
Fayet-Iliopoulos term that triggers the breaking, generating a scale
$\xi $, slightly below the string scale \cite{DSW}. In previous works, it has been demonstrated that models of
this class can naturally explain the values of many low energy
parameters \cite{Ib}. Supersymmetry breaking at the same time still remains a
mystery, so we can not say a lot about an important part of the
phenomenological predictions that these models are capable of
producing, the soft supersymmetry breaking
parameters like squark and gaugino masses. 
These, contribute to processes on which there exist strict experimental
constraints.
For example, they
contribute to fcnc effects, known to be very small. 
A viable model therefore, must have a supersymmetry
breaking mechanism that yields squark masses compatible with low
energy data on these processes. 
One class of models attempting to explain the suppression of the 
supersymmetric contribution to fcnc is decoupling of the third generation by making it much heavier 
than the first two -approximately degenerate in mass \cite{NW}. Another is
quark-squark alignment, in which case 
even if supersymmetry breaking doesn't yield degenerate or decoupled
squarks, 
fcnc is automatically suppressed \cite{LNS}.
A third is a supersymmetry breaking mechanism that yields degenerate (or almost degenerate)
squark masses. Some time ago, a class of models in which supersymmetry breaking is communicated to 
the low energy world by an anomalous $U(1)$ was
proposed, in the context of global supersymmetry \cite{BD}, \cite{HDM}. 
It was soon realized that if the breaking is mainly 
due to nonzero vevs of $D$-terms the resulting soft masses are proportional 
to the anomalous  $U(1)$ charges of the fields and
the desired degeneracy is not achieved. This problem could be overcome if the
anomalous $U(1)$ is family blind, but then
the family structure of the quark and lepton sector would be 
trivial, unless there are additional family dependent, non-anomalous $U(1)$ factors ($Y^{(i)}$)
in the gauge group. We will show below, that the vevs of the $D$-terms 
associated to the non-anomalous $U(1)$'s are proportional to that of the 
anomalous $U(1)$, so they are all contribute the same order of magnitude to 
flavor changing processes and the problem we had before, remains. 
The soft masses, in the context of 
these models, have two types of contributions:
\be m_{{\phi}_i}^2=m_0^2+m_i^2={1\over 12}K_2|<F_S>|^2\;\;\;-
\sum_{a=X,Y^{(1)},...}{q^{a}_{\phi_i}}<D_a> \label{eq:soft} , \ee
where $m_0$ is the (family independent)\footnote{We assume that the
non-universal couplings of the dilaton are suppressed.} contribution
from the dilaton $F$-term, $m_i$ the 
(generically family dependent) contribution from $D$-terms and $K_2$ is the second
derivative of the K$\ddot{\rm a}$hler potential $K$, with respect to 
the real part of the dilaton field $S$: $y={(1/2)}(S+{\overline S})$.
$K$ is an unknown function of $y$.
We will try to 
answer the following question: Is it possible to construct a model in which 
the presence of the family dependent $U(1)$'s is not disasterous for 
flavor physics? We distinguish two 
possible phenomenologically viable scenarios.
\begin{itemize} 
\item $m_0^2<<m_i^2<<m_{1/2}^2$. 
\par The first is a ``no scale'' type of 
scenario, where $m_0$ is very close
to zero and the non-universal $m_i$ are larger but still small
enough, so that when extrapolated to the MSSM scale, they do not give
dangerous contributions to fcnc because the running of the soft masses
to low scale is dominated by (large) gaugino masses. 
Such a boundary condition is obtained when the dilaton is stabilized
at a very small value of $K_2$. This can be achieved by assuming a
weakly coupled form for $K$. An example of such a 
K$\ddot{\rm a}$hler potential was proposed in \cite{HDM}:
\be K=-\ln{(2y)}-{s_0\over y}+{(b+4s_0^2)\over 24y^2},\ee
which as long as $b>0$ and $b\le 1/m^2$, has a minimum near
$y_0=s_0-(1/m)$
and the values of the derivatives at the minimum are given by:
\be K_1=-{1\over 3s_0}; \;\;\; K_2={1\over m^2s_0^4}; \;\;\;
K_3=-{2\over ms_0^4}.\label{eq:Kvac} \ee
Indeed, $|K_1|<1$ and $K_2<<1$,
since for reasonable values of the parameters,
$m$ is a number $\sim 20-80$ (see eq 1.8 and 2.28 below) and
therefore $s_0\sim 3/2$. 
This type of
models however, as we will see, are plagued by charge/color breaking minima
because of the absence of large contribution to $m_0$ from the dilaton, that tends to
stabilize the vacuum. In order, therefore, to construct viable models of
this type, we will have to assume the existence of additional family
independent $F$-term contributions from other moduli that stabilize
the low energy vacuum. 
\item $m_i^2<<m_0^2\simeq m_{1/2}^2$.
\par The second is a 
``minimal sugra'' type of scenario, where in order to suppress fcnc, 
we $require$ that all the $D$-term contributions are very small. We will now
argue that in the extreme case where these exactly vanish, we can make 
predictions for the soft parameters.  
Following \cite{HDM}, upon integrating out the heavy gauge field
associated with $X$ and taking the $D$ term part of its equation of
motion at the minimum, we obtain a relation  
between the vacuum expectation values of the anomalous 
$D$-term $D_X$ and that of the dilaton $F$-term $F_S$:
\be <D_X>=-{1\over 4}|<F_S>|^2\Bigl[{K_3\over K_1}-\bigl({K_2\over
K_1}\bigr)^2\Bigr]\big(1-{16\pi^2\delta_{GS}\over 4}{K_2\over K_1}\bigr)^{-1} ,\label{eq:DFs} \ee 
In the presence of additional non-anomalous $U(1)$ factors, we can similarly
integrate out their heavy gauge fields and equation (\ref{eq:DFs})
still holds \footnote{We assume that there is no appreciable kinetic
mixing between the $U(1)$'s.}. The scale of the FI term can be evaluated from
\be{\xi \over M}={1\over 2}\sqrt{16\pi^2\delta _{GS}K_1}\label{eq:FI}.\ee
For a confining gauge group in the hidden sector, there is a non perturbative
contribution to the superpotential that is of the form
\be W^{(np)}=B e^{-mS}\label{eq:np},\ee
where $m$ is a model dependent (group theoretical) number and the
prefactor $B$ has units of mass cubed.  
The contribution of the dilaton to the scalar potential then becomes:
\be V^S={1\over 4}K_2M^2|F_S|^2={1\over K_2^*M^2}|{W^{(np)}}_S|^2=4m^2B^2
{{e^{-2my}\over K_2M^2}},\ee
where we have used (\ref{eq:np}) and that 
\be F_S={-4\over K_2M^2}{\partial {W^{(np)}}^*\over \partial S}\label{eq:FS}.\ee 
In a dilaton dominated scenario (where $V\sim V^S$), 
the above term dominates the minimization condition $V_1=0$ and therefore 
at the minimum we get the condition
\be {K_3\over K_2}=-2m \label{eq:con1}. \ee
Substituting (\ref{eq:con1}) into (\ref{eq:DFs}), we deduce that in order the
$D$-term contribution to the soft masses to vanish, 
the following has to be satisfied at the minimum:
\be {K_2\over K_1}=-2m\label{eq:con2} .\ee
This implies that in order to have degenerate squarks 
after supersymmetry breaking, the form of the K$\ddot{\rm a}$hler 
potential (at the minimum) has to be of the form:
\be K=ce^{-2my_0}.\ee
The constant $c$ can be fixed from (\ref{eq:FI}). Doing so, we obtain 
for $K_2$ at the minimum:
\be K_2=-{8m\over {16\pi^2\delta_{GS}}}{\bigl({\xi \over M}\bigr)^2}\label{eq:K2}.\ee
Knowing $K_2$, allows us to compute the soft masses from
(\ref{eq:soft}), in a model with known superpotential.
\end{itemize}
We emphasize that the two types of limits are quite different. In the
first ``no scale'' limit, we will use a perturbative form for the 
K$\ddot{\rm a}$hler potential which stabilizes the dilaton at a very
small value of $K_2$ \cite{HDM}. In the ``minimal sugra'' limit, we do not
assume a specific form for $K$, but instead we guess its value at the
minimum, requiring that the $D$-term vanishes. The whole function $K$
is unknown in this case and it may or may not contain both
perturbative and non-perturbative
contributions. One has to be very careful with trying to guess the
form of the whole function. In fact, by taking for $K$ the sum of the
perturbative $-\log{(2y)}$
term and a typical non-perturbative term, makes the vanishing of the
$D$-term rather difficult. This could mean that it is not correct to
impose such a constraint or that the form of $K$ can not naively be guessed.        
The dilaton in this limit, is $assumed$ to be stabilized to $y_0$ and
if it happens, it
happens at a much higher value of $K_2$ than in the ``no scale'' case. 
\par Many models with a 
family dependent, anomalous $U(1)$ constructed in the past, have success explaining the mass ratios 
and mixings in the quark and charged lepton sectors \cite{Ib}. 
Recently however, it was argued that in order a model of this type to 
naturally relate vacuum stability with the see-saw mechanism 
and R-parity conservation, the anomalous $U(1)$ has to
be family blind \cite{BILR}. Therefore, a model that can explain the mass 
hierarchies and mixings has to contain in addition
to the anomalous, other, non-anomalous $U(1)$ factors that are family
dependent. Interestingly, this is 
precisely what happens in most realistic superstring
comactifications \cite{faraggi}. 
Such a model was presented in \cite{ILR}, 
with one anomalous, family blind ($X$) and two non-anomalous family
dependent ($Y^{(1)}$, $Y^{(2)}$) $U(1)$'s. 
The vacuum of this model was 
shown to be stable and free of flat directions associated with any 
invariant of the gauge group \cite{IL}. It reproduced all quark and charged
lepton masses and 
mixings and predicted neutrino masses and mixings compatible with the
solar and 
atmospheric neutrino data. Proton decay 
was within the experimental bounds and R-parity was conserved,
yielding a stable LSP. 
In addition, the model had a hidden sector 
capable of breaking supersymmetry via gaugino condensation. 
We will use the same observable sector as in \cite{ILR} but 
assume a different, simpler hidden sector. In section 2 we show how supersymmetry 
is broken in a general model with many $U(1)$'s and give expressions for
the the soft parameters. In section 3 we apply the general formalism
to an explicit model. In section 4 we give our conclusions. 

%%%%%%%%%%%%%%%%%%%%%%%%%%%%%%%%%%%%%%%%%
\section{Supersymmetry Breaking with $U(1)$'s}
%%%%%%%%%%%%%%%%%%%%%%%%%%%%%%%%%%%%%%%%%

In this section we extend the supersymmetric breaking mechanism of
\cite{BD} and  
\cite{HDM} for the case of one anomalous
and an arbitrary  number of non-anomalous $U(1)$'s. 
The $U(1)$'s break slightly below the string
scale by the vevs of a set of singlet fields that we call $\theta
_k$. The number of these  singlets is equal to the number of the
additional $U(1)$'s, so that their charges form a nonsingular square matrix:
\be{\cal A}=\pmatrix{x_1&x_2&x_3&.\;\;.\;\;. \cr
y^{(1)}_1&y^{(1)}_2&y^{(1)}_3&.\;\;.\;\;. \cr
y^{(2)}_1&y^{(2)}_2&y^{(2)}_3&.\;\;.\;\;. \cr  . & .& .&  .\;\;.\;\;.\cr 
.& .& .&   .\;\;.\;\;.\cr .&  .& .&   .\;\;.\;\;.\cr },\ee 
where the first row contains the charges of 
$\theta _k$ with respect
to the anomalous symmetry $X$, the second row the charges with respect
to the non-anomalous $Y^{(1)}$ and so forth.  
The supersymmetric vacuum is defined to be the solution of the equations
\be D_{X,Y^{(a)}}=0.\ee
Denoting the vevs $<\theta _k>$ by $v_k$, the $D$-flatness condition 
\be {\cal A}\pmatrix {|\theta _1|^2\cr |\theta _2|^2\cr |\theta
_3|^2\cr . \cr .\cr .\cr}=\pmatrix {\xi ^2 \cr 0\cr 0\cr. \cr .\cr .\cr }\ee
and the gauge invariance condition for the mass term of a field $t$
\be {1\over 2}Mt^2\Bigl({\theta _1\over M}\Bigr)^{p_1}\Bigl({\theta _2\over
M}\Bigr)^{p_2}\Bigl({\theta _3\over M}\Bigr)^{p_3}\; . \; . \; .\ee
in the superpotential: 
\be {\cal A}\pmatrix {p_1 \cr p_2\cr p_3\cr . \cr .\cr .\cr }=\pmatrix {n \cr 0\cr 0\cr . \cr .\cr .\cr },\ee
give the supersymmetric vacuum constraint
\be \pmatrix {v^2_1 \cr v^2_2\cr v^2_3\cr . \cr .\cr .\cr }={1\over
\rho} \pmatrix {p_1 \cr p_2\cr p_3\cr . \cr .\cr .\cr } \label{eq:suva}, \ee
where $\xi ^2$ is the Fayet-Iliopoulos term generated by the breaking
of the anomalous $U(1)$, 
and 
$\rho \equiv -n/\xi^2$, where $n$ is 
the $X$ charge of the field $t^2$. $M$ is the cut-off scale
of our theory $\sim 10^{16-17}GeV$ 
\footnote{For the sake of simplicity, we decided to use only one mass
scale in our model, the scale at which the observable
sector gauge couplings unify, even though in certain cases this might
not be the most appropriate.}. 
From (\ref{eq:suva}), we can see that the ratio $p_i/v_i^2=const.$
We will be looking for 
a supersymmetry breaking vacuum in the vicinity of this vacuum.

\par In the following, we assume for simplicity that $G^h$ is a
semi-simple, compact, non-Abelian gauge group and that there is only 
one type of hidden condensates. If there
are other hidden fields besides those forming the condensates, they are singlets of $G^h$. We also
assume that the number of hidden colors $N_c$, is greater than the
number of hidden families $N_f$,
in which case the non-perturbative superpotential is particularly simple.  
Gaugino condensation occurs at a scale where the hidden sector beta function blows up. 
This scale, is calculated from the
renormalization group equation to be
\be \Lambda =Me^{-8\pi ^2k _h(2S)/b_0} =
Me^{-{{{nN_f}\over {{\delta_{GS}}}}}{S\over b_0}}\label{eq:scale} ,\ee
where $k_h$ is the Kac Moody level of the hidden group $G^h$ and $b_0$
is the one loop beta function of the hidden sector. 
Below this scale, condensates  of the hidden ``quark'' fields ${\bf q}_i$ will be formed:
\be t_i=(2{\bf q}_i{\bf \overline q}_i)^{1/2},\ee
where the index $i$ counts the number of hidden families $N_f$. 
In the following we will always assume that it is possible to diagonalize the condensate's mass matrix
and in addition that all the condensates have the same mass. In this case $t$ becomes a diagonal matrix 
with equal entries along the diagonal so we can simplify the
calculation by minimizing the scalar potential for a single 
$t$ and keeping in mind that it is multiplied by an $N_f\times N_f$ unit matrix. 
\par We are ready now to write down the scalar 
potential to be minimized. It's general form is 
\be V=V^0+V^S ,\ee 
where 
\be V^0=\sum _k {\Bigl|{{\partial W}\over {\partial \theta
_k}}\Bigr|^2}+
{\Bigl|{{\partial W}\over {\partial t}}\Bigr|^2}+
\sum _{X,Y^{(1)},Y^{(2)},...}{{1\over {2g^2}}D^2}\;\; ;\ee
the $D$-terms in the above are 
\be D_X=-g_X^2\bigl[x_1|\theta _1|^2+x_2|\theta _2|^2+x_3|\theta _3|^2+...+
{1\over 2}n|t|^2+\xi ^2\bigr]+\; . \; . \; .\ee
\be D_{Y^{(a)}}=-g_a^2\bigl[y^{(a)}_1|\theta _1|^2+y^{(a)}_2|\theta _2|^2+
y^{(a)}_3|\theta _3|^2+... \bigr]+\; . \; . \; . 
\footnote{The dots in the two expressions stand for contributions
from all other fields. These however should not be allowed to take
vevs for obvious reason.}
\;\; ,\ee
where $a$ runs over only the non-anomalous $U(1)$'s.  
The superpotential is given by 
\be W=W^{(p)}+W^{(np)}={1\over 2}Mt^2\Bigl({\theta _1\over M}\Bigr)^{p_1}
\Bigl({\theta _2\over M}\Bigr)^{p_2}\Bigl({\theta _3\over
M}\Bigr)^{p_3}...
+\bigl({{d_a\over {2d_r}}-N_f}\bigr)
\Bigl({2\Lambda ^{b_0\over 2}\over t^2}\Bigr)^{1\over {{d_a\over {2d_r}}-N_f}}\label{eq:w} \label{eq:W},\ee
where $d_r$ is the Dynkin index of the representation $r$ of the hidden
gauge group ($r=a$ is the adjoint).
Using (\ref{eq:scale}) and (\ref{eq:w}), we can express the 
model dependent constant $m$ in terms of group theoretical numbers:
\be m={{8\pi ^2k_h}\over {{d_a\over {2d_r}}-N_f}}.\ee
Consider now the minimization conditions 
\be \theta _1 {\partial V\over {\partial \theta _1}}=(p_1-1)|F_{\theta
_1}|^2+p_1|F_{\theta _2}|^2+p_1|F_{\theta _3}|^2+p_1{2\over t}F_t^*W^{(p)}-|\theta
_1|^2(x_1D_X+y^{(1)}_1D_{Y_1}++y^{(2)}_1D_{Y_2}+...)=0 \label{eq:min1}\ee
\be t {\partial V\over {\partial t}}=2|F_{\theta }|^2-|F_t|^2+{4\over
t}F_t^*(W^{(p)}+
{1\over {({d_a\over {2d_r}}-N_f)^2}}W^{(np)})
-{1\over 2}nt^*D_X+t|{V^S}_t|=0 \label{eq:min2} \ee
where $|F_{\theta }|^2=|F_{\theta _1}|^2+|F_{\theta _2}|^2+|F_{\theta _3}|^2+...$  and 
\be t{V^S}_t={-2\over {d_a\over {2d_r}}-N_f}{V^S} \;\; {\rm where}
\;\; {V^S}_t=
{\partial V^S\over \partial t}.\ee
Defining $N_c\equiv {2d_a\over d_r}$,
(which for fields transforming in the fundamental of $SU(N_c)$ is just the usual color $N_c$),
the $F$-terms entering the above equations are
\be F^*_{\theta _k}=-{\partial W\over \partial \theta _k}=-{p_k\over \theta _k}W^{(p)}\ee
and 
\be  F^*_{t}=-{\partial W\over \partial t}=-{2\over t}(W^{(p)}-{1\over {N_c-N_f}}W^{(np)}).\ee
We are looking for a minimum in the vicinity of the DSW \cite{DSW} vacuum:
\be <F_t>\sim 0,\;\;\;\; <t>\sim 0 \;\;\;{\rm and }\;\;\; <\theta _k>\sim \xi .\ee
The first of the above conditions, implies that
\be W^{(p)}={1\over {N_c-N_f}}W^{(np)} \ee
in the vacuum.
Then, (\ref{eq:min2}) becomes
\be {2\over t}F_t^*W^{(p)}=-ft|{V^S}_t|-f|F_{\theta }|^2,\ee
where we have introduced $f=(N_c-N_f)/(N_c-N_f+1)$.
Substituting this into (\ref{eq:min1}), we get
\be (p_1-1)|F_{\theta _1}|^2+p_1(1-f)(|F_{\theta }|^2-|F_{\theta
_1}|^2)-
p_1f|F_{\theta }|^2-p_1ft{{|V^S}_t|\over 2}=|\theta _1|^2
(x_1D_X+y_1^{(1)}D_{Y^{(1)}}+y_1^{(2)}D_{Y^{(2)}}+...)\ee
which in the vacuum (where $<\theta _a>=v_a$ and everything 
is evaluated at the minimum), becomes
\be {W^{(p)}}^2\rho ^2[(1-f)(p_1+p_2+p_3+...)-1]-\rho f t{|{V^S}_t|\over
2}=x_1<D_X>+y_1^{(1)}<D_{Y^{(1)}}>+
y_1^{(2)}<D_{Y^{(2)}}>+\; ...,\ee 
where we have used (\ref{eq:suva}).
Notice that the left hand side of the above equation does not depend
on the index $''1''$, so all the minimization conditions with respect
to all $\theta _k$ can be obtained from this, by interchanging the
subscripts of the right hand side by $''k''$. This in turn implies
that we can solve these equations     
for the vevs of the $D$-terms:
\be \pmatrix {<D_X>\cr <D_{Y^{(1)}}>\cr <D_{Y^{(2)}}>\cr .\cr .\cr
.\cr }=
{\bf C}({\cal A}^{-1})^{T}\pmatrix {1\cr 1\cr 1\cr .\cr .\cr .\cr}\ee
where 

\begin{eqnarray} {\bf C}={W^{(p)}}^2\rho ^2\Bigl[{1\over {N_c+1-N_f}}(p_1+p_2+p_3+...)-1\Bigr]-
\rho {{N_c-N_f}\over {N_c+1-N_f}} t{{|V^S}_t|\over 2} \cr
=(\epsilon ^2{\hat m}^2n^2) \Bigl[ \Bigl( {1\over
{N_c+1-N_f}}(p_1+p_2+p_3+...)-1 \Bigr) -
\Bigl( {(8\pi^2 k_h)^2\over {(N_c-N_f+1)}}{1\over {nK_2}}{\bigl({\xi \over
M}\bigr)^2}\Bigr) \Bigr]. 
\cr \label{eq:DC}\end{eqnarray}
The above relation implies that the values of the $D$-terms are proportional:
\be <D_{Y^{(a)}}>={A_a\over A_X}<D_X>\ee
where $A_X$ is the sum of the entries of the first column of ${\cal
A}^{-1}$, $A_1$ 
is the sum of the entries of the second column of ${\cal A}^{-1}$,
etc.
This shows that in general, the $D$-terms contribute to supersymmetry
breaking, but also that for $det{\cal A}\ne 0$, if $<D_X>$ vanishes,
then all other $<D_{Y^{(a)}}>$ vanish as well.
\par Using (\ref{eq:FS}), (\ref{eq:scale}) and (\ref{eq:W}), we find
that the vacuum value of the $F$-term associated with the dilaton is
\be <F_S>=\epsilon{\hat m}(8\pi ^2k_h){4\over K_2}\Bigl({\xi \over
M}\Bigr) ^2\ee
and we have defined, as usual, the helpful variables
\be {\hat m}=M\lambda _1^{p_1}\lambda _2^{p_2}\lambda _3^{p_3}...
\;\;\; {\rm with} \;\;\; \lambda _k\equiv <{\theta _k\over M}>,\ee
\be 1>>\epsilon ={<t>^2\over 2\xi ^2}=\Bigl({\Lambda \over
\xi}\Bigr)^{b_0\over {2N_c}}\Bigl({\xi
\over {\hat m}}\Bigr)^{1-{N_f\over N_c}}, 
\;\;\; {\rm with} \;\;\; \Lambda =Me^{-8\pi
^2k_h{(2y_0)\over b_0}}.\label{eq:eps} \ee
Here, $y_0=<y>=1/g(M)^2$ where $g(M)$ is the value of the gauge coupling
at the unification scale $M$ and we have assumed that the dilaton gets somehow stabilized to a 
reasonable value $y_0$\footnote{We will see that by reasonable we mean
a value $\sim 1.5$ (see Appendix) }.
The one loop beta function is given by
\be b_0=3d_a-\sum _r{d_r}.\; 
\footnote{In our normalization of the
indices, for $SU(N_c)$ with $N_f$ families of ``quarks'' and
``antiquarks'', the beta function is $b_0=2(3N_c-N_f)$.}\ee
We normalize the Dynkin indices so that 
\be Tr(T^r_aT^r_b)=d_r\delta _{ab}\ee
with $T^r_a$ being the generators of $G^h$ in the representation $r$. 
\par Having the expressions of the vevs of the $D$ and $F$-terms, 
we can now calculate the soft masses. 
We can assign to each field 
-generically denoted by $\phi _i$, with $i$ being a
family index- a set of numbers $n_k$ such that the term 
\be \bigl({\theta _1\over M} \bigr)^{n^i_1}\bigl({\theta _2\over M}  \bigr)^{n^i_2} 
\bigl({\theta _3\over M}  \bigr)^{n^i_3}.\; .\; .\; \phi _i
\label{eq:inv} \;\;\;
\footnote{The invariants of the whole gauge group are just gauge
invariant polynomials of such terms.}\ee
is invariant under the $U(1)$'s. 
\par The soft masses then can be written as
\begin{eqnarray} {\tilde m}_{\phi_i}^2=m_0^2+m_i^2=
\Bigl[ {4 \over {\sqrt{3K_2}}}(\epsilon {\hat m})(8\pi^2 k_h){\bigl({\xi
\over M}\bigr)^2}
\Bigr]^2+\Bigl[\sqrt{(n^i_1+n^i_2+n^i_3+\cdots){\bf C}} \Bigr]^2
.\end{eqnarray}
\par The gaugino masses are
\be m_{1/2}={<F_S>\over {2y_0}}={1\over y_0}\sqrt {3\over K_2}{m}_0.\label{eq:gaugino}\ee
\par The trilinear soft couplings are 
\be A^{[u,d,e]}_{ij}\sim
<F_S>Y^{[u,d,e]}_{ij}=A_0m_{1/2}Y^{[u,d,e]}_{ij}\equiv
{\bf \rm a}_0Y^{[u,d,e]}_{ij},\ee
where $A_0$ is a constant of order of one and $Y^{[u,d,e]}_{ij}$ is the 
corresponding Yukawa coupling in the superpotential. 
We now consider the two different boundary conditions. 

\begin{itemize}   
\item {\bf $m_0^2<<m_i^2<<m_{1/2}^2$}
\par First, notice that in (\ref{eq:DC}), the second term inside the brackets dominates
over the first for reasonable values of the parameters, so the first
term can safely be neglected. Then, the ratios 
that are expected to be small in this limit are
\be {m_0^2\over m_i^2}=-{4\over 3}{\bigl({N_c-N_f+1\over
n}\bigr)}{{(\xi /M)^2}\over {(n^i_1+n^i_2+n^i_3\cdots
)}}\label{eq:ratMar} \ee
and 
\be {m_i^2\over m_{1/2}^2}={y_0^2K_2 \over 3}{m_i^2\over m_{0}^2}.\ee
In order both these ratios be simoultaneously suppressed, $K_2$ has to be
rather small, as we argued in the introduction.

\item {\bf $m_{1/2}^2\simeq m_0^2>>m_i^2$}
\par We saw that
if the conditions (\ref{eq:con1}) and (\ref{eq:con2}) are
satisfied then the only contribution to the soft masses comes 
from the dilaton $F$-term:
\be {m}_{0}= {4 \over {\sqrt{3K_2}}}(\epsilon {\hat m})(8\pi^2 k_h){\bigl({\xi
\over M}\bigr)^2} \label{eq:squarks} \ee
with $K_2$ given by (\ref{eq:K2}) and it is manifestly flavor and family universal for all fermions. 
The common gaugino mass and the trilinear couplings are
\be {\bf a}_0\simeq m_{1/2}={1\over y_0}\sqrt{3\over K_2}m_0.\ee
\end{itemize}

%%%%%%%%%%%%%%%%%%%%%%%%%%%%%%%%%%%%%%%
\section{The Model}
%%%%%%%%%%%%%%%%%%%%%%%%%%%%%%%%%%%%%%%

We now apply the general formalism of the previous section to an
explicit model. As we mentioned before, we will use the visible
sector of \cite{ILR} and introduce a slightly different hidden sector. For a detailed
discussion of the phenomenological consequences related purely to the visible sector we 
refer the reader there. For completeness, we give the Yukawa matrices of
the visible sector in the Appendix. 
\par The fields present in the visible sector are
\par $\;\;\; {\bf 1.}$ Three singlets of the non-Abelian part of the
gauge group that take 
vevs and uniquely break the $U(1)$'s (called $\theta _i$). This sector
is necessarily anomalous.
\par $\;\;\; {\bf 2.}$ Three chiral families in the ${\bf 27 (16+10+1)}$ of $E_6 (SO(10))$ except the singlet  
of $SO(10)$. 
\par $\;\;\; {\bf 3.}$ One standard-model vector like pair of Higgs weak doublets. It
turns out that this is a model with $\tan {\beta}$ is of order of one (see Appendix). 
\par $\;\;\; {\bf 4.}$ Four singlets of the non Abelian part of the gauge group $\Sigma
_k$, that do not take vevs and they are 
introduced to cancel the anomalies of the $\theta _i$ fields. 
\par The hidden sector has three ($N_f=3$) families of vector-like fields 
${\bf q}_i$ and ${\bf \overline q}_i$,
transforming as ${\bf N_c}$ and ${\bf {\overline N}_c}$ of $G_h$, with ${\bf
N_c}$ 
being the fundamental representation of $G_h$.
There are, in addition, a set of $G_h$ singlet hidden fields $T_j$, $j=1,...,N_T$. 

\par The gauge group is
\be SU(3)^c\times SU(2)_W\times U(1)_Y\times U(1)_X\times
U(1)_{Y^{(1)}}\times 
U(1)_{Y^{(2)}}\times D_1 \times G^{h}\ee
where $D_1$ is a discrete symmetry acting on the visible vector-like fields in the ${\bf 10}$ of $SO(10)$, as 
\be E_i\rightarrow -E_i,\;\;\; {\overline E}_i\rightarrow -{\overline
E}_i,\;\;\; {\bf D}_i\rightarrow 
-{\bf D}_i,\;\;\; {\bf \overline 
D}_i\rightarrow -{\bf \overline D}_i\ee
and 
\be G^h\equiv SU(N_c), \;\; {\rm with}\;\; N_c>N_f.\ee

We denote by $V$ and $V'$ the non-anomalous $U(1)$'s in $E_6$ (besides the
regular hypercharge $Y$),
according to
\be E_6 \subset SO(10)\times U(1)_{V'}\ee
with 
\be {\bf 27}={\bf 16}_1+{\bf 10}_{-2}+{1}_4\ee
where the subscript is the $U(1)$ value. The other $U(1)$ is
\be SO(10)\subset SU(5)\times U(1)_V,\ee
with
\be {\bf 16}={\bf \overline 5}_{-3}+{\bf 10}_{1}+{\bf 1}_{5};\;\; {\rm and} \;\;\;
{\bf 10}=
{\bf \overline 5}_{2}+{\bf 5}_{-2}.\ee
The charges of the visible fields 
in the ${\bf 16}+{\bf 10}$ of $SO(10)$, (${\bf Q}$, ${\bf \overline u}$, 
${\bf \overline d}$, ${L}$, ${\overline e}$, ${\overline N}$, ${E}$,
${\overline E}$, ${\bf D}$, ${\bf \overline D}$),
under the three $U(1)$'s ($X$, $Y^{(1)}$, $Y^{(2)})$ are
\be X=(-1-{3\over 20}V+{1\over 4}V')\pmatrix {1&0&0 \cr 0&1&0 \cr 0&0&1}, \ee
\be Y^{(1)}={1\over 5}(2Y+V)\pmatrix {2&0&0 \cr 0&-1&0 \cr 0&0&-1}, \ee
\be Y^{(2)}={1\over 4}(V+3V')\pmatrix {1&0&0 \cr 0&0&0 \cr 0&0&-1} \ee
and hypercharge is normalized so that the triangle anomaly
coefficient with one anomalous gauge field and two hypercharge (and
weak) gauge bosons obey the relation:
\be \bigl(XYY\bigr)={5\over 3}\bigl(XSU(2)SU(2)\bigr).\ee
In the following, we will call $\bigl(XSU(2)SU(2)\bigr)\equiv C_W$ for simplicity. 
\par The charges of the rest of the visible sector singlet fields are 
\be \theta _i:\;\; {\cal A}=\pmatrix {1&0&0 \cr 0&-1&1 \cr 1&-1&0}\;\;
{\rm and}\;\;\;\; \Sigma _i:\;\; 
\pmatrix {-1/2&-1/2&0&0 \cr 0&0&1/2&-1/2 \cr -9/4&-7/4&9/4&7/4}, \ee
where the first row contains the charges under $X$ and the second (third)
row contain the charges 
under $Y^{(1)}(Y^{(2)})$. The matrix ${\cal A}$ implies that all three
$U(1)$'s break at precisely the same scale. Then, the different
expansion parameters $\lambda _k$, are all equal to the same $\lambda
\equiv \xi /M$.
\par As mentioned before, there are in addition three families of vector like hidden fields transforming under
the fundamental (anti-fundamental) representation ${\bf N_c}$ (${\bf
{\overline N}_c}$) of $SU(N_c)$ with charges under the $U(1)$'s as
%%%%%%%%%%%%%%%%%
\hskip 2cm
\begin{center}
\begin{tabular}{|c|c|c|}
\hline          
$ $ & ${\rm {\bf q}_i}$ & ${\bf \overline q}_i$  \\ \hline \hline   

$X$& $-3$ & $-3$ \\

$Y^{(1)}$ & $2/{N_c}$ & $-2/{N_c}$ \\                                                            
 
$Y^{(2)}$ & $1/2$ & $-1/2$ \\ \hline

\end{tabular}
\end{center}
\vskip 0.3cm 
%%%%%%%%%%%%%%%%%%
where $i=1,2,3$ since $N_f=3$. 
This implies that $p_1=p_2=p_3=6$. 
\par The fields $T_j$ have no
charges under the non-anomalous $U(1)$'s and their charge under $X$ is $-3$. This last set of
fields is given for completeness, since their only purpose here is, to adjust the 
gravitational anomaly to be compatible with the Green-Schwarz
mechanism. For a hidden sector with $k_h=1$, we need
\be N_T=45-6N_c.\ee 
\par As it stands, this model is anomaly free. 
The Green-Schwarz relations 
\be {C_j\over k_j}=constant, \;\;\;  {\rm for \;\; all \;\; j}
\label{eq:GS} \ee
are all satisfied with the Kac-Moody levels of the non-Abelian factors all
equal to $1$, including $k_h$.  
$C_j$ in the above are the non-zero anomaly coefficients associated with the different gauge
factors and $k_j$ the corresponding Kac-Moody levels. 
We distinguish again the two limits and present for each case 
examples in the context of this model. 

\begin{itemize}
\item {\bf $m_0^2<<m_i^2<<m_{1/2}^2$}
\par For this, ``no scale'' case, the universal contribution to the soft
masses is very small (it vanishes for all practical purposes: $m_0\sim
0$) and the
family dependent contribution from the $D$-terms is
\be m_{i}^{({\bf Q},{\bf \overline u},{\bf \overline d},{L},{\overline
e})}= \sqrt{\bf C}\cdot \sqrt{{\bf n}_{i}^{({\bf Q},{\bf \overline u},{\bf \overline
d},{L},{\overline e})}},\label{eq:nonun}\ee  
where ${\bf n}_{i}^{\phi }$ is the sum of the exponents defined in
(\ref{eq:inv}) for the field $\phi $ and $i$ is its family index. They
are easily calculable in our model (see Appendix).
The gaugino masses and the trilinear couplings are 
\be x\equiv m_{1/2}\simeq {\bf a_0}.\ee 
In table 1, we show some typical values that our model gives for the parameters 
$\sqrt{\bf C}$ and $x$ with $\xi /M$ as a free
parameter. Unfortunately, all these models have problems associated
with charge and/or color breaking minima of the scalar potential and
therefore can not be considered as viable \cite{CCB} as they stand. This is a
generic feature of the no scale boundary conditions. There is however the 
possibility of other moduli $F$-terms contributing to $m_0$, with  
vevs large enough to protect the vacuum
\footnote{I thank S. Martin for pointing this out.}. In such a case,
we could have a viable model of the ``no scale'' type. 
An example is, if
these contributions for the $N_c=4$ and $\xi/M=0.24$ case (see table
1.), were $\sim
200\;GeV$. The resulting low energy spectrum would then be very
similar to the ``minimal sugra'' spectrum that we present below and we
show in table 2. 

\item {\bf $m_{1/2}\simeq m_0^2>>m_i^2$}
\par As an example, we take $N_c=5$ and $\xi/M=0.28$ 
\footnote{There is nothing particularly deep in the choice of these
parameters. The choice of $N_c=5$ is motivated by the fact that it is
probably the only value that results in reasonable squark masses and
the choice $\xi/M=0.28$ is the value of the expansion parameter
$\lambda _c$ that we found in ref. \cite{ILR}, a number close to
the Cabbibo angle (see Appendix). If we use a different
value for $\xi/M$, $m_0$ and $m_{1/2}$ will change.}
and use as input
$y_0=1.48$ (see Appendix). For these values, the small expansion parameter 
is $\epsilon \simeq 0.86\cdot 10^{-5}$ and 
the condensation scale becomes $\Lambda \simeq 2\cdot 10^{12}\;
GeV$. The dilaton is now stabilized at $K_2\simeq 1.3$ (see
\ref{eq:K2}). 
Substituting these values into (\ref{eq:squarks}) and
(\ref{eq:gaugino}),
we obtain 
\be {m}_{0}\simeq 200\;\; {\rm GeV}\; ; \;\;\; 
{m}_{1/2}\simeq 200\;\; {\rm GeV}\; ; \;\;\; {\bf \rm a_0}\simeq 200\;\; {\rm GeV}.\ee
\par Of course, 
these parameters, are predicted at the unification
scale $M$, so we have to extrapolate 
their values to $M_Z$ to obtain the low energy spectrum. In 
table 2,
we show the MSSM parameters corresponding to this particular model
\cite{Diego}. It is an example of a phenomenologically viable model,
with no charge/color breaking minima,
consistent with EWSB and fcnc.  
Experimental signatures that this type of models could imply is for example
a trileptonic signal 
in $p{\bar p}$ collisions: $p{\bar p}\rightarrow {\tilde
C}_1^{\pm}+{\tilde N}_2+X\rightarrow 
l_1{\bar \nu}_1{\tilde N}_1+l_2{\bar \nu}_2{\tilde N}_2+X$ (three
leptons and missing energy) \cite{AN}, or at the LHC, the decay $h^0\rightarrow
\gamma \gamma$ \cite{Haber}. Also, $e^+e^-$ annihilation could pair produce
the lowest mass charginos and sfermions. 
The exact degeneracy of the squark masses is a result of the
simultaneous conditions (\ref{eq:con1}) and (\ref{eq:con2}). If these 
are not exactly obeyed, then the squark masses split with the
splittings 
proportional to the vevs of the $D$-terms times the $U(1)$ 
charges of the quark fields. 
Soft masses with non-universal contributions from $D$-terms
may be a more realistic scenario but in that case the mass differences,
especially between the first two families should be 
small, in order to avoid conflict with experimental bounds on fcnc. 
\end{itemize}
\par We finally emphasize that using at certain places $M_{string}$ instead of $M\equiv
M_{GUT}$, can alter the numerology quite significantly but    
our purpose here is just to show that one can build a
complete model based on $U(1)$'s, 
which is consistent with present experimental bounds. 

%%%%%%%%%%%%%%%%%%%%%%%%%%%%%
\section{Conclusion}
%%%%%%%%%%%%%%%%%%%%%%%%%%%%%
We extended the dilaton dominated, (global) supersymmetry breaking mechanism by gaugino
condensation in the hidden sector, communicated to the visible sector
by an anomalous $U(1)$, to the case when the gauge group contains
additional non-anomalous $U(1)$'s. 
We saw that our model was capable of producing phenomenology in two
different limits, depending on the choice of the hidden sector and the
K$\ddot{\rm a}$hler potential. One was a ``no scale'' type limit,
where $m_0\sim 0$, $m_i$ small but nonzero and $m_{1/2}={\bf a}_0$ large. 
The dilaton was stabilized at a very low value of
$K_2$ due to a weak coupling choice for $K$. 
We could not find a viable model in this regime without some vacuum
stabilization mechanism. 
The other, was a ``minimal sugra'' type limit.  
In this scenario, we did not stabilize the dilaton,
but by requiring the $D$-term contributions to the soft masses be
zero, we were able to obtain the value of the K$\ddot{\rm a}$hler function and its
derivatives in the vacuum, assuming that the dilaton gets stabilized to
a reasonable value. For a particular choice of parameters, 
we obtained at unification scale
degenerate soft masses and gaugino masses both $\sim 200\;GeV$. We
extrapolated the values of the soft parameters to $M_Z$ and made low
energy numerical
predictions for this example.

\vskip 1cm 
{\bf Acknowledgements}
\vskip .5cm

I would like to thank P. Ramond and P. Bin\'etruy for discussions and
suggestions. I am especially grateful to S. Martin and to D. Casta${\tilde {\rm n}}$o for
their comments and help.   

\section{Appendix}
We review some crucial facts concerning the visible sector of the
model of ref. \cite{ILR}.
\begin{itemize} 
\item The superpotential $W$ contains the Yukawa terms that give after EW
symmetry breaking masses to the quarks, leptons and
neutrinos:
\be W_{Yukawa}=Y^{[u]}_{ij}{\bf Q}_i{\bf \overline u}_jH_u
+Y^{[d]}_{ij}{\bf Q}_i{\bf \overline d}_jH_d+
Y^{[e]}_{ij}{L}_i{\overline e}_jH_d+     
Y^{[\nu]}_{ij}{L}_i{\overline N}_jH_u+
Y^{[0]}_{ij}M{\overline N}_i{\overline N}_j+\cdots .\ee
The Yukawa matrices that give the suppression to the operators are
result of the breaking of the $U(1)$'s. They are parametrized by
$\lambda _c $, the Cabbibo angle. In the model under consideration
they turn out to be in the quark sector:
\be Y^{[u]}=\pmatrix {\lambda _c^8 & \lambda _c^5 & \lambda _c^3 \cr
\lambda _c^7 & \lambda _c^4 & \lambda _c^2 \cr
\lambda _c^5 & \lambda _c^2 & 1  },\;\;\;\;
Y^{[d]}=\lambda _c^3\pmatrix {\lambda _c^4 & \lambda _c^3 & \lambda _c^3 \cr
\lambda _c^3 & \lambda _c^2 & \lambda _c^2 \cr
\lambda _c & 1 & 1  }\ee
and in the lepton sector:
\be Y^{[e]}=\lambda _c^3\pmatrix {\lambda _c^4 & \lambda _c^5 & \lambda _c^3 \cr
0 & \lambda _c^2 & 1 \cr
0 & \lambda _c^2 & 1  }, \;\;\; 
Y^{[\nu]}=\pmatrix {\lambda _c^8 & \lambda _c^7 & \lambda _c^3 \cr
\lambda _c^5 & \lambda _c^4 & 1 \cr
\lambda _c^5 & \lambda _c^4 & 1  },\;\;\;
Y^{[0]}=\lambda _c^7 \pmatrix {\lambda _c^6 & \lambda _c^5 & \lambda _c \cr
\lambda _c^5 & \lambda _c^4 & 1 \cr
\lambda _c & 1 & 0  }.\ee
One consequence of these matrices is that $\tan{\beta }$ is of order
of one, as
one can read off from the $(33)$ elements of $Y^{[u]}$ and $Y^{[d]}$. 
\item The vector-like matter in the ${\bf 10}$ of
$SO(10)$ enters $W$ with mass terms:
\be W_{VL}=MY^{[E]}_{ij}{\overline E}_i{E}_j+
MY^{[D]}_{ij}{\bf \overline D}_i{\bf D}_j+\cdots, \ee
where
\be Y^{[E]}=\pmatrix {\lambda _c^3 & \lambda _c^7 & \lambda _c^{11} \cr
\lambda _c^5 & \lambda _c^9 & \lambda _c^{13} \cr
\lambda _c^7 & \lambda _c^{11} & \lambda _c^{15}  }\;\; {\rm and} \;\;
Y^{[D]}=\pmatrix {\lambda _c^3 & \lambda _c^7 & \lambda _c^{9} \cr
\lambda _c^5 & \lambda _c^9 & \lambda _c^{11} \cr
\lambda _c^9 & \lambda _c^{13} & \lambda _c^{15}  }.\ee  
The mass eigenvalues of both fields are
\be \{\lambda _c^3M,\lambda _c^9M,\lambda _c^{15}M\}.\ee
\item The powers appearing in (\ref{eq:nonun}), are
\be {\bf n}_i^{{\bf Q}}= 
\Bigl[({9\over 10}+{19\over 10}+{37\over 30}),
({9\over 10}+{9\over 10}+{37\over 30}),
({9\over 10}-{1\over 10}+{7\over 30})\Bigr]\ee
\be {\bf n}_i^{{\bf \overline u}}= 
\Bigl[({9\over 10}+{19\over 10}+{77\over 30}),
({9\over 10}+{9\over 10}+{17\over 30}),
({9\over 10}-{1\over 10}-{13\over 30})\Bigr]\ee
\be {\bf n}_i^{{\bf \overline d}}= 
\Bigl[({3\over 10}+{3\over 10}+{29\over 30}),
({3\over 10}+{3\over 10}-{1\over 30}),
({3\over 10}+{3\over 10}-{1\over 30})\Bigr]\ee
\be {\bf n}_i^{{L}}= 
\Bigl[({3\over 10}+{3\over 10}+{23\over 10}),
({3\over 10}+{3\over 10}-{7\over 10}),
({3\over 10}+{3\over 10}-{7\over 30})\Bigr]\ee
\be {\bf n}_i^{{\overline e}}= 
\Bigl[({9\over 10}+{19\over 10}-{1\over 10}),
({9\over 10}+{9\over 10}+{19\over 10}),
({9\over 10}-{1\over 10}+{9\over 10})\Bigr]\ee
\be {\bf n}_i^{{H_u}}= 
\bigl[({-9\over 5}+{1\over 5}+{1\over 5})\bigr]\ee
\be {\bf n}_i^{{H_d}}= 
\bigl[({9\over 5}-{1\over 5}-{1\over 5})\bigr]\ee
\item To calculate the unification scale and the value of the gauge coupling
at unification, we have to take into account the three new thresholds
due to vector-like matter. By doing so, we find 
\be {1\over g(M)^2}=1.48\ee
and $M=3.4\cdot 10^{16}\; GeV$. This is the $M$ that is used consistently as our
only mass parameter, even though we are aware that this might not be
the most appropriate in some cases. 
\item The value of the parameter $\xi/M$ is undetermined, as long as
the K$\ddot{\rm a}$hler potential is unknown. Since however we use it
as the expansion parameter in the mass matrices, it is 
implicitly assumed to be a number
close to the Cabbibo angle ($\sim 0.22$). Assuming, for example, for $K$ its usual
tree level form $K=-\log{(2y)}$ and using $M$ and $g(M)$ from above, we
obtain $\xi/M=0.28$.   
\item The $\mu $ term has zero charge under all $U(1)$'s. It could
therefore appear in $W$ and in the K$\ddot{\rm a}$hler potential, but
since one does not get pure mass terms in a string spectrum, we assume
its presence only in the K$\ddot{\rm a}$hler potential. Therefore, 
after supersymmetry breaking, at
low energy, a $\mu $ term of the correct order (few hundred $GeV$) will be generated by
the Guidice-Masiero mechanism. 
\item There is no kinetic mixing between the $U(1)$'s.
\item R-parity is exactly conserved.
\end{itemize}

%%%%%%%%%%%%%%%%%%%%%%%%%%%%%%%%%%%%%%%%%%%%%%%%%%%%%%%%%%%%%%%%%%%%%%%%%

%%%%%%%%%%%%%%%%%%%%%%%%%%%%%%%%%%%%%%%%%%%%%%%%%%%%%%%%%%%%%%%%%%%%%%%%%

%%%%%%%%%%%%%%%%%%%%%%%%%%%%%%%%%%%%%%%%%%%%%%%%%%%%%%%%%%%%%%%%%%%%
\eject
%%%%%%%%%%%%%%%%%%%%%%%%%%%%%%%%%%%%%%%%%%%%%%%%%%%%%%%%%%%%%%%%%%%%%

\begin{table}

\caption{Values for $x$ and $\sqrt{\bf C}$ at the high scale $M$, for different choices of
$N_c$ and $\xi/M$, in the ``no scale'' regime. }

\vskip 0.5cm
\begin{center}
\begin{tabular}{|c|c|c|c||c|c|c|c|}

           \hline  & & & & & & &    \\
$N_c=5 $ & ${\xi/M }$ & $x(GeV)$ & ${\sqrt{\bf C}}(GeV)$ &
$N_c=4 $ & ${\xi/M }$ & $x(GeV)$ & ${\sqrt{\bf C}}(GeV)$
                                                            \\  & & & & & & &   \\
                       \hline \hline 
& & & & & & &    \\${}$ & ${0.10 }$ & $60  $ & $4  $  & ${}$ & ${0.22}$ & $60$ & $1$ \\     
& & & & & & &    \\${}$ & ${0.12 }$ & $435 $ & $22 $  & ${}$ & ${0.24}$ & $190$ & $3$  \\   
& & & & & & &    \\${}$ & ${0.125}$ & $680 $ & $33 $  & ${}$ & ${0.26}$ & $565$ & $8$ \\   
& & & & & & &    \\${}$ & ${0.13 }$ & $1035$ & $48 $  & ${}$ & ${0.28}$ & $1533$ & $20$ \\   

                                                \hline

\end{tabular}
\end{center}
\end{table}

%%%%%%%%%%%%%%%%%%%%%%%%%%%%%%%%%%%%%%%%%%%%%%%%%%%%%%%%%%%%%%%%%%%%%%%%%%%%%%%%%%

\begin{table}

\caption{The first column contains the name of the parameter. 
The second column contains the low energy value of the parameter for a ``minimal
sugra'' type model corresponding to $N_c=5$, $\xi/M=0.28$ and
$y_0=1.48$.
The input to the RGE's at $M$, is $m_0=m_{1/2}={\bf \rm a}_0=200\;$
GeV, $\tan{\beta}(M_Z)=4$, $sgn(\mu )=+1$ and $m_t=175\;$ GeV.
The values of $({\tilde B},{\tilde W},{\tilde g})$ are their lowest
order pole masses.}

\vskip 0.5cm
\begin{center}
\begin{tabular}{|c|c|}

           \hline  &    \\
{\bf Parameter} & {\bf Value at $M_Z$ ($GeV$ if a mass)}  

                                                            \\  &    \\
                       \hline \hline 
&    \\$({M_Z},{v_{Higgs}})$ & $(90.4,174.1)$ \\  \hline   
&    \\$\tan{\beta}$ & $4$ \\  \hline 
&    \\$({\tilde B},{\tilde W},{\tilde g})$ & 
$(62,122,367)$ \\  \hline 
&    \\$({\alpha_3},{\alpha_2},{\alpha_1},{\sin{\theta_W}})$ & $(0.116,0.033,0.0165,0.232)$ \\  \hline 
&    \\$({Y^{[u]}_{33}},{Y^{[d]}_{33}},{Y^{[e]}_{33}})$ & $(1,0.08,0.042)$ \\  \hline 
&    \\$(B,\mu)$ & $(125,233)$ \\  \hline 
&    \\$({\bf \tilde u}_L,{\bf \tilde u}_R,{\bf \tilde d}_L,{\bf
\tilde d}_R,{\tilde e}_L,{\tilde e}_R,{\tilde {{\nu}_e}}_L) $ &
$(407,399,414,394,238,216,226)$ \\  \hline 
&    \\$({\bf \tilde c}_L,{\bf \tilde c}_R,{\bf \tilde s}_L,{\bf
\tilde s}_R,{\tilde \mu}_L,{\tilde \mu}_R,{\tilde {{\nu}_{\mu}}}_L) $ & 
$(407,399,414,394,238,216,226)$ \\  \hline 
&    \\$({\bf \tilde t}_L,{\bf \tilde t}_R,{\bf \tilde b}_L,{\bf
\tilde b}_R,{\tilde \tau}_L,{\tilde \tau}_R,{\tilde {{\nu}_{\tau}}}_L) $ & 
$(435,283,395,364,238,215,226)$ \\  \hline 
&    \\$(h^0,H^0,A^0,H^{\pm})$ & $(104,346,343,352)$ \\  \hline 
&    \\$({\tilde C}_1,{\tilde C}_2)$ & $(95,270)$ \\  \hline 
&    \\$({{\tilde N}_1},{{\tilde N}_2},{{\tilde N}_3},{{\tilde N}_4})$ & $(53,99,-240,272)$ \\  \hline 
&    \\$LSP\rightarrow \;({\tilde N}_1)$ & $53$ \\  \hline 

                                                \hline

\end{tabular}
\end{center}
\end{table}

%%%%%%%%%%%%%%%%%%%%%%%%%%%%%%%%%%%%%%%%%%%%%%%%%%%%%%%%%%%%%%%%%%%%%%%%%%%%%%%%%%

%%%%%%%%%%%%%%%%%%%%%%%%%%%%%%%%%%%%%%%%%%%%%%%%%%%%%%%%%%%%%%%%%%%%%%%%%
\end{document}